# Simple, self-assembling, single-site model amphiphile for water-free simulation of lyotropic phases

Somajit Dey and Jayashree Saha

Department of Physics, University of Calcutta, 92, A.P.C Road, Kolkata-700009, India, Email: dey.somajit@gmail.com

Computationally, low-resolution coarse-grained models provide the most viable means for simulating the large length and time scales associated with mesoscopic phenomena. Moreover, since lyotropic phases in solution may contain high solvent to amphiphile ratio, implicit solvent models are appropriate for many purposes. By modifying the well-known Gay-Berne potential with an imposed uni-directionality and a longer range, we have come to a simple single-site model amphiphile that can rapidly self-assemble to give diverse lyotropic phases without the explicit incorporation of solvent particles. The model represents a tuneable packing parameter that manifests in the spontaneous curvature of amphiphile aggregates. Apart from large scale simulations (e.g. the study of self-assembly, amphiphile mixing, domain formation etc.) this novel, non-specific model may be useful for suggestive pilot projects with modest computational resources. No such self-assembling, single-site amphiphile model has been reported previously in the literature to the best of our knowledge.

## 1 Introduction

Surfactants or "surface active agents" are so called because of their ability to reduce the surface tension of liquids. Macromolecules that act as surfactants in water generally consist of a hydrophilic (water-loving) head linked to one or more hydrophobic (water-hating) chains (tails). Since the hydrophobic part has an affinity for oil and the hydrophilic part an affinity for water, these molecules are also called amphiphiles.

Phenomena associated with water solution of amphiphiles span many scales in length and time. At extreme dilution the amphiphiles get adsorbed at the water surface or stay dispersed as monomers. Above a certain temperature (Krafft point) and a certain concentration (critical micelle concentration) single chain surfactants aggregate into micelles [1, 2, 3, 4, 5]. Normally micelles are not shape persistent and their aggregation number (number of surfactants in a micelle) may vary [6]. Spherical, elliptical and cylindrical micelles have been observed [6, 7]. With increasing concentration, the amphiphiles are known to generate a variety of ordered aggregation phases depending on the type of amphiphiles. These lyotropic mesophases include tubular middle phase, neat phase from lamellar bilayers, cubic phase, sponge phase, multiply-connected bilayers etc. [6, 8, 9].

The importance of lyotropic liquid crystals (mesophases in solution) cannot be overestimated in biological, therapeutic and industrial context. Analysis of the relevant processes are, however, very challenging due to the complexity of the interactions and the fact that many scales are involved simultaneously. Computer simulation, therefore, holds a very special place in the study of amphiphile aggregates. Since atomistic computer simulations employ the most detailed and chemically specific models, they are unable to probe, within viable processor time, the largest length and time scales associated with events like self-assembly into aggregates, amphiphile mixing, fusion of aggregates etc. [10]. Moreover, complete atomistic detail may actually obscure the fundamental mechanisms underlying these processes providing no insight into them. We, therefore, need models and simulation techniques with inherent time and length scales not too small compared to the scales to be probed. The standard approach in this direction is the use of coarse-grained (CG) models. Coarsening of models by elimination of some of the degrees of freedom, smoothes the phase-space energy surface and speeds up dynamics by allowing larger time steps [10]. It also reduces the number of site-site interactions for a given number of macromolecules implying a larger length scale. However, the small size of CG water molecules compared to the amphiphiles, combined with their number, limit the scales of a simulation. Most CG water models act merely as mediators of an effective hydrophobic bonding [11, 12]. Implicit solvent (IS) amphiphile models do away with the CG water by mimicking this hydrophobic interaction with some specialised interparticle force-field instead [10]. An ISCG model thereby speeds up computation manifold as it concentrates only on the amphiphiles and hence, offers a means to look into the largest time and length scales associated with amphiphile aggregates. Self-assembly into thermally stable aggregates is, however, a great challenge for ISCG models and consequently only a few such models are available in the literature that show successful unassisted self-assembly [13, 14]. These models typically constitute the amphiphiles from a number of beads of different species (polar or hydrophilic, apolar or hydrophobic, linker) with flexible inter-bead bonds [10, 15, 16, 17]. Although the beads interact amongst themselves by simple force laws, the models are invariably multi-site. Coarsening further, the need for a single-site, self-assembling ISCG model amphiphile with a simple force-field was, therefore, hugely felt. Such a model is reported here presently.

The rest of this paper is laid out as follows. In section 2, the motivations guiding our model design are presented. Section 3 then gives the model. Section 4 discusses how the model is meant to satisfy the driving motivations. Section 5 reports some of the phases obtained with molecular dynamics (MD) simulation with this model.

## 2 Motivations

### 2.1 Packing parameter

A most influential concept regarding amphiphile assembly is that of packing parameter [18, 19]. This incorporates a thermodynamic-geometric model whereby an amphiphile in stable aggregates is represented as an oblong molecule shaped just so that such molecules can closely fit together into an aggregate of the correct curvature. Hydrophobic chains in surfactant molecules may exist in a multitude of conformations varying from the fully extended all-trans state to curled gauche states in the oily bulk of the lyotropic mesophases. Packing parameter, $P$ is a ratio between a "steric area", $a_s$, in connection to this incompressible hydrophobic bulk and an "average equilibrium area", $a_e$, at the aggregate-water interface [19]. Let $V_{hp}$ denote the average hydrophobic volume per amphiphile in a close packed aggregate. If $l_{hp}$ be an appropriate weighted average of hydrophobic chain length over all relevant conformations then, $a_s = V_{hp}/l_{hp}$. Therefore, $P = V_{hp}/(a_e l_{hp})$ [19]. FIG.1 and 2 illustrate the essential ideas. $a_e$ depends upon many factors such as headgroup size, hydration of the headgroup, presence of salt in solution, protonation, hydrophobic tail length etc. [20, 19]. $V_{hp}$ depends on the number of hydrophobic chains, saturation of chains, thermal chain conformation distribution etc. $l_{hp}$ for alkanes is, to a good approximation, 80% of the extended chain length [19].

## 2.2 Hydration forces

Water molecules are more localised in the neighbourhood of the hydrophilic groups. Any displacement of amphiphile aggregates, therefore, perturbs hundreds of water molecules that then resist the aggregate motion with energy on the scale of the aggregates. These hydration forces prevent aggregates from coming into contact. Hydrophilic group identity and methylation, hydrophobic chain conformation and heterogeneity all may affect the hydration of surfactants. Hydration force is generally considered to decay exponentially with distance between the aggregates with distance scale of 1-3 Å [21].

## 2.3 Self-assembly

Free energy gain in transfer of hydrophobic chains from water to oily bulk is the main driving force for surfactant self-assembly [22, 23, 24, 25]. Formation of the water-hydrocarbon interface and headgroup repulsion, however, acts against it [25, 26]. Loss of entropy due to orientational confinement in aggregates also acts as a limiting cause [25]. The net effect can be modelled with directed amphiphiles interacting through short range forces favouring certain relative orientations to others. FIG.3 shows five of these orientations in order of decreasing preference from left to right with arrows directed towards the headgroup from the tail of the amphiphile. Phenomenology suggests that the side-side parallel configuration of two surfactants, *a*, should be more favourable than the corresponding anti-parallel configuration, *c*, as flipping of amphiphiles is a relatively rare phenomenon in amphiphile aggregates. End-end anti-parallel configuration with hydrophilic heads away from each other, *b*, is similarly favourable compared to *c*. However, hydrophobic interaction and lipophilic (oil-loving) cohesion favours *a* over *b*. Hydration pressure, repelling hydrated headgroups in vicinity of each other, makes the end-end headgroup facing configuration *e* unfavourable compared to *c*. Hydrophobic and lipophilic interactions also suggest that the end-end parallel configuration *d* should be favourable compared to *e* but unfavourable compared to *c*.

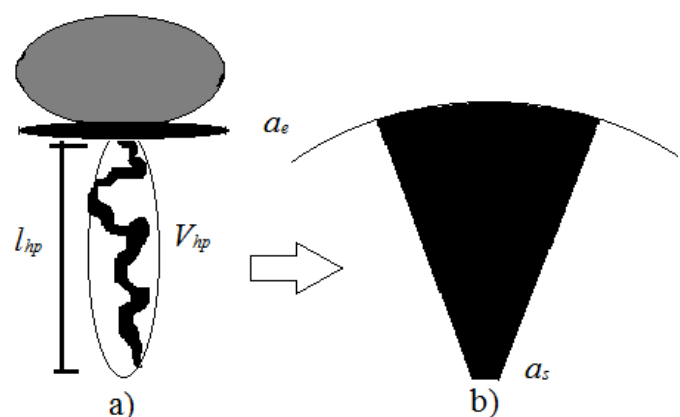

**FIG.1: Packing parameter illustration with $P<1$. a) Hydrated hydrophilic part in grey, equilibrium headgroup area $a_e$ as black disc, single hydrophobic tail as black chain, chain enclosing ellipsoid volume $V_{hp}$ and length $l_{hp}$. b) Equivalent oblong profile in black with top surface area $a_e$ and bottom surface area $a_s$. This profile fits perfectly into a close packed aggregate with positive surface curvature as in micelles.**

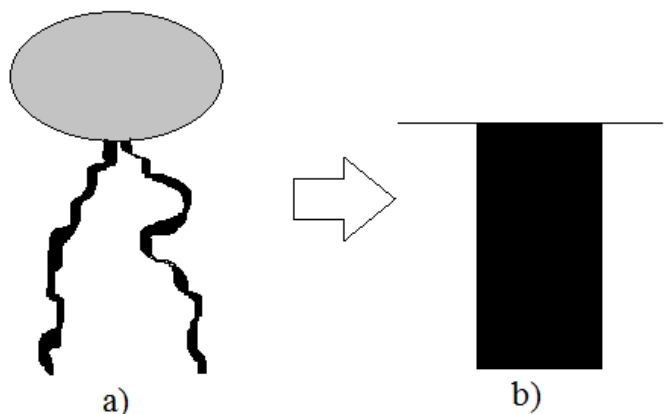

**FIG.2: Packing parameter illustration with $P=1$. Shades bear the same meaning as in FIG.1. a) Surfactant with double hydrophobic chains. b) The oblong profile fits perfectly into a close packed aggregate with zero curvature as in lamellar bilayers.**

From previous modelling efforts it is known that Lennard-Jones (LJ) like forces with typical $r^{-7}$ attractive tail ($r = $ distance) are unable to show unassisted self-assembly into stable fluid aggregates in simulations. Interactions with somewhat longer range are commonly deemed necessary for solvent-free self-assembly **[14]**.

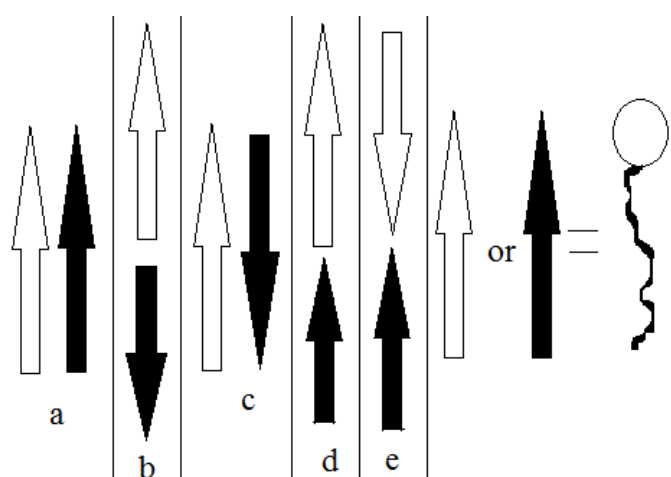

**FIG.3: Relative orientations of amphiphiles in order of decreasing preference from *a* to *e*. Directed arrow significance also shown with a single chain surfactant cartoon. Light and dark shades used for purposes of clarity.**

## 3 Model

Our model amphiphile consists of a soft-core, directed prolate spheroid with a tuneable anisotropic force-field. Spheroid geometry appears most prominently in the popular Gay-Berne (GB) force-field and since our model is inspired by it having some terms in common, we first present the GB model **[27]**.

### 3.1 Gay-Berne force field

Taking two identical spheroids $i$ and $j$ with centres at $\mathbf{r}_i$ and $\mathbf{r}_j$ and major (or minor) axis along unit vectors $\hat{\mathbf{u}}_i$ and $\hat{\mathbf{u}}_j$, the GB ellipsoidal contact distance is given by

$$\sigma_{GB}(\hat{\mathbf{r}},\hat{\mathbf{u}}_i,\hat{\mathbf{u}}_j) = \sigma_0[1-\frac{1}{2}\chi\{\frac{(\hat{\mathbf{r}}.\hat{\mathbf{u}}_i+\hat{\mathbf{r}}.\hat{\mathbf{u}}_j)^2}{1+\chi(\hat{\mathbf{u}}_i.\hat{\mathbf{u}}_j)}+\frac{(\hat{\mathbf{r}}.\hat{\mathbf{u}}_i-\hat{\mathbf{r}}.\hat{\mathbf{u}}_j)^2}{1-\chi(\hat{\mathbf{u}}_i.\hat{\mathbf{u}}_j)}\}]^{-\frac{1}{2}}, \quad (1)$$

where $\hat{\mathbf{r}} = (\mathbf{r}_i - \mathbf{r}_j)/r$, $r$ being the centre-centre distance. Above, the anisotropy parameter $\chi$ is given by $\{(\sigma_e/\sigma_s)^2-1\}/\{(\sigma_e/\sigma_s)^2+1\}$ where $\sigma_e$ ($\sigma_s$) means the end-end (side-side) contact distance. $\sigma_{GB}$, scaled by $\sigma_0$, determines the steric profile of the GB ellipsoid. The overall GB potential is given as

$$V_{GB}(r,\hat{\mathbf{r}},\hat{\mathbf{u}}_i,\hat{\mathbf{u}}_j) = 4\varepsilon_0 \varepsilon^\nu \varepsilon_{GB}^\mu (R^{-12}-R^{-6}), \quad (2)$$

where $R = (r-\sigma_{GB}+\sigma_0)/\sigma_0$ and $\varepsilon = \{1-\chi^2(\hat{\mathbf{u}}_i.\hat{\mathbf{u}}_j)^2\}^{-1/2}$, $\varepsilon_0,\nu,\mu$ are parameters. The term $\varepsilon_{GB}$ refers to an energy ellipsoid and has formal similarity with $\sigma_{GB}$ as

$$\varepsilon_{GB}(\hat{\mathbf{r}},\hat{\mathbf{u}}_i,\hat{\mathbf{u}}_j) = 1-\frac{1}{2}\chi'\{\frac{(\hat{\mathbf{r}}.\hat{\mathbf{u}}_i+\hat{\mathbf{r}}.\hat{\mathbf{u}}_j)^2}{1+\chi'(\hat{\mathbf{u}}_i.\hat{\mathbf{u}}_j)}+\frac{(\hat{\mathbf{r}}.\hat{\mathbf{u}}_i-\hat{\mathbf{r}}.\hat{\mathbf{u}}_j)^2}{1-\chi'(\hat{\mathbf{u}}_i.\hat{\mathbf{u}}_j)}\}, \quad (3)$$

$\chi'$ being related to the anisotropy in well-depth.

### 3.2 The amphiphile model

FIG.4 depicts the model we envisage as a directed spheroid against a realistic cartoon of a single chain and a double-chain surfactant. $\hat{\mathbf{u}}$ is now taken to concur with this directionality. With a GB like ellipsoidal core, our model potential then reads,

$$V(r,\hat{\mathbf{r}},\hat{\mathbf{u}}_i,\hat{\mathbf{u}}_j) = 4\varepsilon_0(R^{-12}-R^{-6}+1/4)-\varepsilon_{wd}(\hat{\mathbf{r}},\hat{\mathbf{u}}_i,\hat{\mathbf{u}}_j), \text{ if } r < r_l$$
$$= s(r)\varepsilon_{wd}(\hat{\mathbf{r}},\hat{\mathbf{u}}_i,\hat{\mathbf{u}}_j), \text{ if } r_l \leq r < r_u \quad (4)$$
$$= 0, \text{ otherwise.}$$

Above, $r_l = (2^{1/6}-1)\sigma_0+\sigma_{GB}$ and $r_u = r_l + \text{range}$, where range signifies the tuneable range of interaction. $s(r)$ is a cubic switching function given by $s(r) = (r_u-r)^2(3r_l-2r-r_u)/\text{range}^3$. The all essential anisotropic well-depth function $\varepsilon_{wd}(\hat{\mathbf{r}},\hat{\mathbf{u}}_i,\hat{\mathbf{u}}_j)$ is given as

$$\varepsilon_{wd} = \varepsilon_0[\nu_1(\hat{\mathbf{u}}_i.\hat{\mathbf{u}}_j)+\nu_2(\hat{\mathbf{u}}_i.\hat{\mathbf{r}}-\hat{\mathbf{u}}_j.\hat{\mathbf{r}})-\nu_3(\hat{\mathbf{u}}_i.\hat{\mathbf{r}})(\hat{\mathbf{u}}_j.\hat{\mathbf{r}})+1]^{\nu_0}\varepsilon_{GB}(\hat{\mathbf{r}},\hat{\mathbf{u}}_i,\hat{\mathbf{u}}_j) \quad (5)$$

$\nu_0,\nu_1,\nu_2,\nu_3$ are four parameters to be input externally along with the parameter $\chi'$ associated with $\varepsilon_{GB}(\hat{\mathbf{r}},\hat{\mathbf{u}}_i,\hat{\mathbf{u}}_j)$. The next section

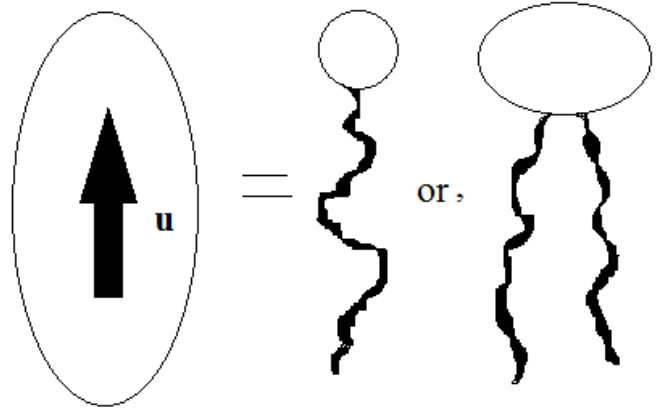

**FIG.4:** Our model ISCG amphiphile as a directed spheroid with anisotropic interactions interpreted as single or double chain surfactants with hydrated headgroup.

discusses this and the choice of parameters in more detail.

## 4 Interpretations

For any $\hat{\mathbf{r}},\hat{\mathbf{u}}_i,\hat{\mathbf{u}}_j$ the generic $r$ dependence of the model potential is shown in FIG.5. As is readily seen, the attractive tail falls more slowly compared to the LJ type potentials hence fulfilling the need for a longer range (Sec. 2.3). The variability of range here should offer a tuneable thermal stability of the amphiphile aggregates as in the other models in literature with tuneable range **[15]**.

Orientational preferences, as in FIG.3, are achieved in our model via differences in well-depths for different orientations. For example, inter-amphiphile repulsions in configurations d and e in FIG.3 are modelled with negative well-depths (FIG.6). A familiar interaction where relative orientations of two directed vectors are energetically distinguished is the dipole-dipole interaction. The first and third term within square brackets in equation (5) are actually inspired by similar terms in the dipole-dipole interaction. It may also be noted that the repulsive tail in FIG.6 can be approximated as an exponential decay reminiscent of the distance dependence of the hydration forces (Sec. 2.2).

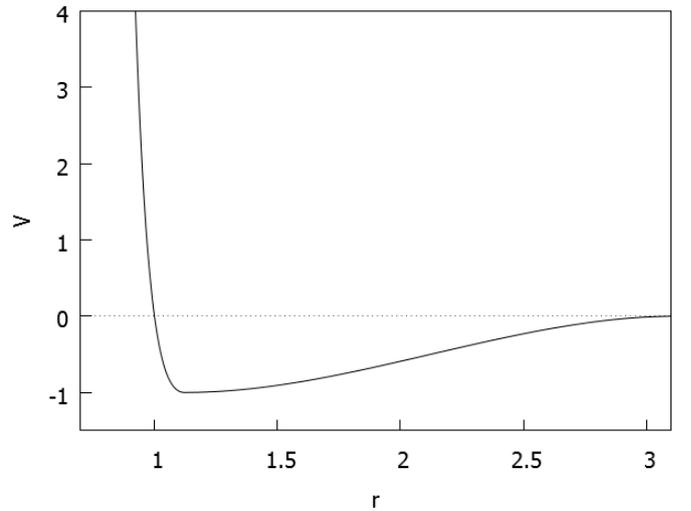

**FIG.5:** $V(r)$ for some choice of $\hat{\mathbf{r}},\hat{\mathbf{u}}_i,\hat{\mathbf{u}}_j$ such that $\sigma_{GB}=1$ and $\varepsilon_{wd}=1$. $\text{range}=2$.

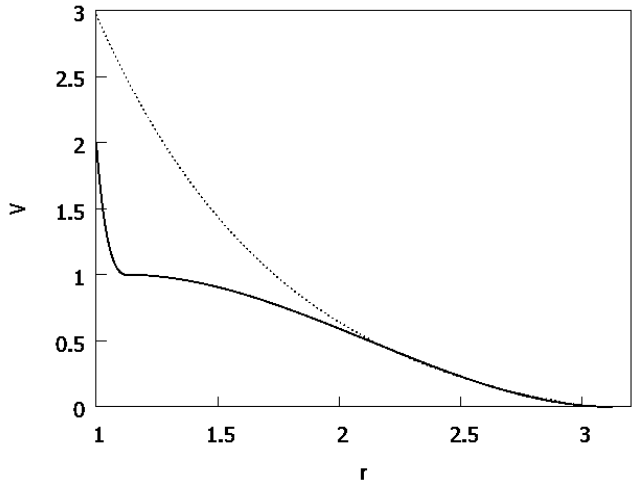

**FIG.6:** Everything same as in FIG.5 except $\varepsilon_{wd} = -1$. The dotted line depicts an exponential fitted to the tail part.

Well-depth for the cross (X) configuration of the two spheroids is $\varepsilon_0$ from equation (5). For the end-end configuration $\varepsilon_{GB} = (1 - \chi')/(1 + \chi')$ $(= \varepsilon_e, \text{say})$ as obtained from equation (3). Hence we can also use $\varepsilon_e$ instead of $\chi'$ as a parameter. If the well-depth for the $a$ and $c$ configurations (FIG.3) are given as $wd_a$ and $wd_c$ then from equation (5),

$$wd_a = \varepsilon_0 (v_1 + 1)^{v_0} \quad (6)$$

$$wd_c = \varepsilon_0 (1 - v_1)^{v_0} \quad (7)$$

Solving these gives $v_0, v_1$ (It is, however, necessary that $v_0$ be odd as otherwise the negative well-depths required to produce the above-mentioned repulsions will never be generated). Similarly, the knowledge of well-depths corresponding to the remaining three configurations in FIG.3 simultaneously give $v_2, v_3$ and $\chi'$ (or equivalently $\varepsilon_e$). In other words, the five parameters in the well-depth function (equation (5)), therefore, are able to completely reproduce the order (and degree) of preference of the simplest characteristic model configurations (FIG.3). If we take

$$v_0 > 0, \quad (8)$$

then from equations (6) and (7), and the order of preference $wd_a > wd_c$, we have the inequality

$$\begin{aligned} v_1 + 1 > 1 - v_1 > 0 \\ \text{or,} \quad 1 > v_1 > 0. \end{aligned} \quad (9)$$

Proceeding similarly with the other configurations in FIG.3, in the given order of preference, we find

$$v_1 + v_2 > v_3, \quad (10)$$

$$v_1 < v_3 - 1. \quad (11)$$

The condition $wd_a > wd_b > wd_c$ gives upper and lower bounds on $\varepsilon_e$ as

$$\left( \frac{1 + v_1}{1 - v_1 + 2v_2 + v_3} \right)^{v_0} > \varepsilon_e > \left( \frac{1 - v_1}{1 - v_1 + 2v_2 + v_3} \right)^{v_0}. \quad (12)$$

The inequalities (8)-(12) serve in checking the consistency of any simple choice of the parameters.

As we proceed from the most preferred configuration, $a$, in FIG.3 to the next preferred orientation, $b$, by gradually increasing the angle $\theta$ (FIG.7) between the spheroids, $\hat{\mathbf{u}}_i \cdot \hat{\mathbf{u}}_j$ and $\varepsilon_{GB}$ decrease while $\hat{\mathbf{u}}_i \cdot \hat{\mathbf{r}} - \hat{\mathbf{u}}_j \cdot \hat{\mathbf{r}}$ and $-(\hat{\mathbf{u}}_i \cdot \hat{\mathbf{r}})(\hat{\mathbf{u}}_j \cdot \hat{\mathbf{r}})$ increase. This implies a crossover with a maximum $\varepsilon_{wd}$ (equation (5)) at some $\theta$, say $\theta_{max}$, between 0 and $\pi$. This orientation is preferred even to configuration $a$ in FIG.3, and is actually the most preferred orientation (FIG. 8). As evident from FIG.7, this minimum energy configuration can be taken to represent a packing parameter model (Sec. 2.1, FIG.1, 2) with interim angle $\theta_{max}$ between its long sides. For $\theta_{max} \neq 0$, therefore, a non-zero curvature is expected in the aggregates. Other model parameters remaining constant, $\theta_{max}$ increases with increasing $\varepsilon_e$. Hence, $\varepsilon_e$ (or $\chi'$) can be regarded as the parameter governing phase curvature or packing parameter.

In view of the above discussion, our generic single-site model amphiphile can be interpreted as any surfactant with a polar head and apolar tail(s) treating both the hydrated headgroup and the hydrophobic and lipophilic tail interactions implicitly. It may be noted that mixtures of amphiphiles of different species can easily be modelled by choosing parameters of the inter-species interaction differently from the intra-species ones. However, all species must have the same length as interaction of spheroids of two different lengths has no interpretation in our model.

## 5 Computer simulations

In the absence of any explicit aqueous phase for pressure coupling, NVT molecular dynamics (MD) was performed instead of NPT with the above model for systems of identical amphiphiles **[28, 29]**. It may, however, be noted that for sufficiently homogeneous systems and periodic bilayers appropriate barostatting may be designed **[30]**. The Nose-Hoover (NH) algorithm was employed for the canonical thermostatting, carefully avoiding the Toda oscillation as far as possible **[31, 32, 33]**. Sufficiently large time steps were used without compromising the desired conservation of an appropriate quantity that remains conserved in NH MD **[33]**. A simple integrator for linear molecules was used to rotate the uniaxial spheroids **[34]**. Microcanonical NVE simulations were also undertaken that gave results consistent with the NVT ones **[28, 29]**. Here, however, only the NVT results will be reported. The relevant expressions for the forces and torques are given in the Appendix. Periodic boundary conditions and a fixed cubic simulation box were employed for all the simulations. System sizes (number of amphiphiles) chosen were 256, 500, 1372 and 6912. All simulations were performed on a desktop computer.

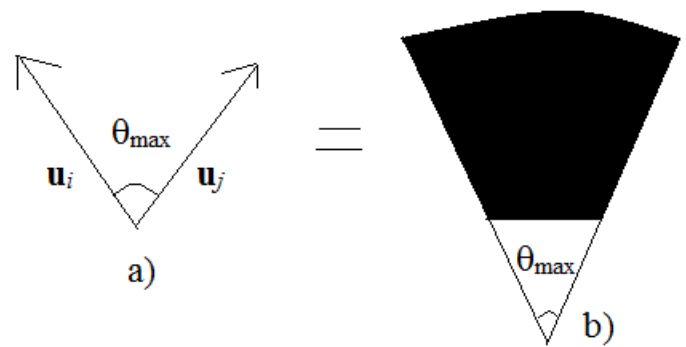

**FIG.7:** a) Angle $\theta_{max}$ between two directed spheroids with directed axes $\hat{\mathbf{u}}_i$ and $\hat{\mathbf{u}}_j$. b) Equivalent packing parameter model.

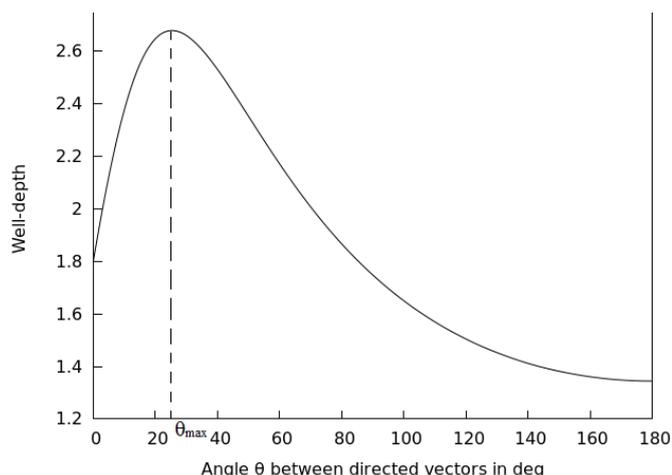

FIG.8: Well-depth, as a function of $\theta$ for configurations as in FIG.7a). Choice of parameters: $v_0 = 1, v_1 = 0.8, v_2 = 4, v_3 = 3, \varepsilon_e = 0.12$.

Self-assembly from randomised initial configurations was found to be quite rapid with aggregates forming within 3000 (micelles) to 4000 (bilayers) steps. This is at par with some established multi-site ISCG models **[13, 14]**. Increasing $\varepsilon_e$ while retaining rest of the parameters in our model as constants, phases with more and more positive curvature were obtained. This trend was consistently reproduced for varying parameter choices and system sizes. For a given choice of parameters, however, system size affects the density at which a specific phase is to be formed.

Stable bilayers and micelles were obtained for a range of $\varepsilon_e$ depending on the temperature. Stability was checked by complete disassembly through heating followed by cooling to the temperature under study. The phases, if regained, would most likely be stable.

Under suitable conditions box-spanning stable bilayers forming neat phase **[6]** were obtained (FIG.9). Otherwise, many bilayer patches remained in the simulation box. Sometimes multiply connected ramp structures were observed. Although no sealed vesicles were obtained, curved bilayers reminiscent of vesicular cross-sections were observed at large system sizes (of the order of 1000 particles in the simulation box). Configurations like pores, passages and necks were also observed at these system sizes [35]. At lower temperatures and low values of $\varepsilon_e$, crystal-like gel phases were observed with long range ordering as opposed to the fluid bilayer phases (no long-range ordering) at the other end of the spectrum **[15]**. FIG.10 shows that the root-mean-square displacement from the equilibrium stable bilayer initial configuration in FIG.9 increases with MD steps. Sans long range order, the bilayer in FIG.9 is thus seen to be fluid indeed **[28]**. With increasing length of the model spheroid, interdigitation became prominent in the (gel like) smectic A **[6]** configuration.

At very high $\varepsilon_e$, spherical micelles were obtained (FIG.11). Micelles were also found to arrange themselves in a cubic crystal as in an isotropic phase [6]. For $\varepsilon_e$ with values interim between the bilayers and spherical micelles, cylindrical and ellipsoidal micelles were found often in coexistence with spherical micelles (FIG.12). Micelles were generally seen to be thermally more stable compared to bilayers.

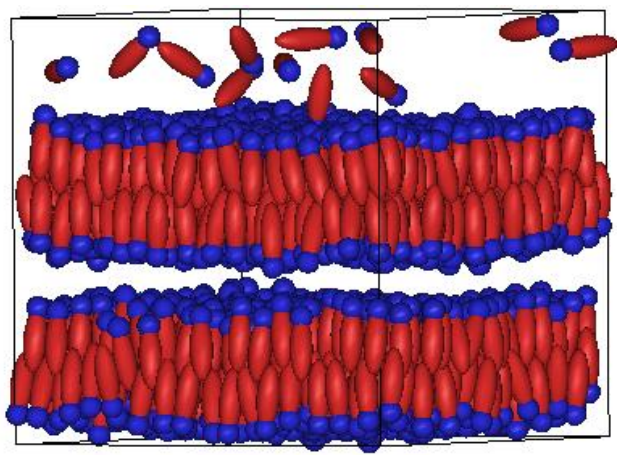

FIG.9: Stable bilayer; 1372 **particles in cubic simulation box with periodic boundary;** $v_0 = 1, v_1 = 0.8, v_2 = 4, v_3 = 3, \varepsilon_e = 0.052$, range=3 ; amphilphile moment of inertia = 4, length = 3, temperature = 2.5 **in reduced units. Dark spheres signify headgroups for directed surfactants.**

## 6 Conclusions

A simple, single-site, coarse grained model amphiphile has been presented for implicit solvent simulations of lyotropic phases. The model is able to show rapid unassisted self-assembly into aggregates with diverse morphology from micelles to bilayers. Inspired by Gay-Berne mesogens it consists of a uniaxial directed spheroidal core. Effective hydrophobicity and hydration pressure is mimicked by varying well-depths inspired by terms in the well-known dipole-dipole interaction. The model amphiphile bears close resemblance to the packing parameter concept prevalent in the field of lyotropic aggregation.

*In silico* simulations with this model were also reported. Different lyotropic phases were successfully generated for varying choices of model parameters. A study of mixtures of different amphiphiles modelled with different such parameter sets is currently being pursued by the present authors.

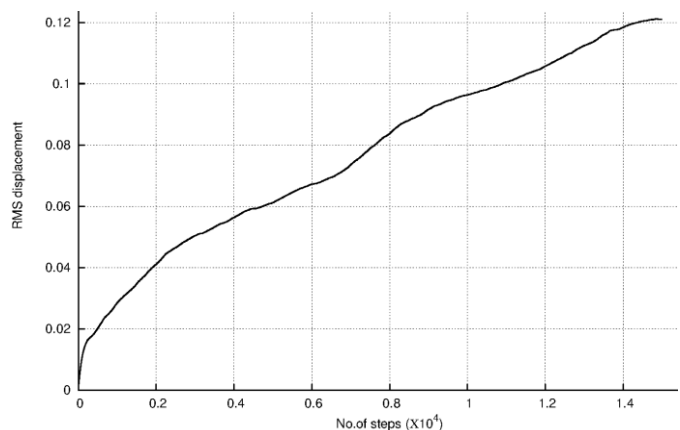

FIG.10: Post equilibration RMS displacement for 15000 steps from the configuration in FIG. 9

| 5

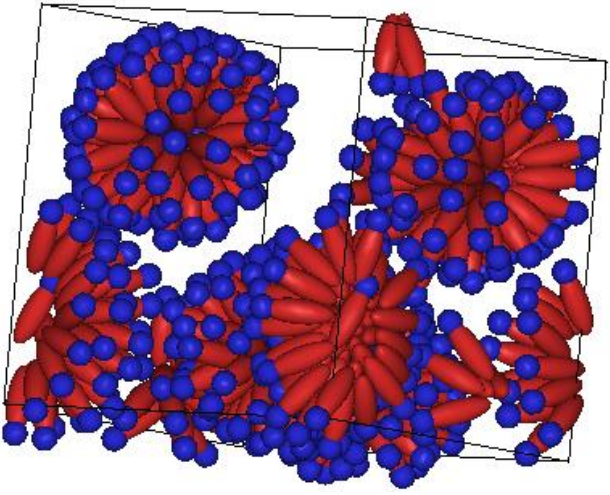

**FIG.11: Micelles.** Model parameters are same as in FIG.9 except $\varepsilon_e = 0.15$ **and** temperature=2.0 **here. System size = 500 particles within cubic simulation box. Dark spheres signify headgroups for directed surfactants.**

Self-assembling ISCG amphiphiles currently available in the literature broadly fall into two classes. Some invoke a realistic molecular architecture **[30]** while others use a single chain of beads **[15, 16, 17]**. Some models are intended to reproduce general elastic or mesoscopic properties of amphiphile aggregates while others are calibrated to reproduce more microscopic features like area per amphiphile in lamellar bilayers under vanishing surface tension **[13]**. Ours is a non-specific qualitative model that shows unassisted self-assembly into diverse lyotropic phases and hence is suitable for generic mesoscopic studies. Other such non-specific models are invariably multi-site making explicit use of polar and apolar beads to represent hydrophilic and hydrophobic portions respectively. Ours is unique in its use of a simple single-site anisotropic interaction with imposed directionality for implicit treatment of both the hydrated polar headgroup and the hydrophobic interaction. No such self-assembling, single-site amphiphile model has been reported previously in the literature to the best of our knowledge.

The present model responds to a long-standing need for an efficient single-site model in context of simulating large scale phenomena emerging from aggregation of nano-scale molecules. Such phenomena span cellular biology, lyotropic solution based industries, and pure soft matter physics. Studies of these lyotropic phases are scarce compared to the thermotropic liquid crystals due to computational complexities. Simple single-site anisotropic interaction models like Gay-Berne ellipsoids have contributed significantly in the thermotropic studies. With the present model the same is hoped in the field of lyotropic liquid crystals.

## Acknowledgements

One of the authors (S.D) is grateful to CSIR for financial assistance in terms of a Junior Research Fellowship under File No. 09/028(0960)/2015-EMR-I. Simulations were run on computers provided under the UPE-UGC scheme.

## Appendix: expressions for forces and torques

The force exerted on amphiphile $i$ by amphiphile $j$ is $\mathbf{F}_{ij} = -\vec{\nabla}_\mathbf{r} V$ and that on $j$ by $i$ is $\mathbf{F}_{ji} = -\mathbf{F}_{ij}$. Now,

$$\vec{\nabla}_\mathbf{r} V = \frac{\partial V}{\partial r}\hat{\mathbf{r}} + \frac{1}{r}[\vec{\nabla}_{\hat{\mathbf{r}}}V - \{(\vec{\nabla}_{\hat{\mathbf{r}}}V).\hat{\mathbf{r}}\}\hat{\mathbf{r}}]$$
$$= \left\{\frac{\partial V}{\partial r} - \frac{(\vec{\nabla}_{\hat{\mathbf{r}}}V).\hat{\mathbf{r}}}{r}\right\}\hat{\mathbf{r}} + \frac{1}{r}\vec{\nabla}_{\hat{\mathbf{r}}}V. \quad (13)$$

The torque on $i$ due to $j$ is given by $\mathbf{T}_{ij} = -\hat{\mathbf{u}}_i \times \vec{\nabla}_{\hat{\mathbf{u}}_i}V$ and that due to $i$ on $j$, by $\mathbf{T}_{ji} = -\hat{\mathbf{u}}_j \times \vec{\nabla}_{\hat{\mathbf{u}}_j}V$. However, $\mathbf{F}_{ij}$, $\mathbf{T}_{ij}$ and $\mathbf{T}_{ji}$ are related by the vector equation, $\mathbf{T}_{ji} + \mathbf{T}_{ij} + \mathbf{r} \times \mathbf{F}_{ij} = 0$. This follows directly from the rotational invariance of $V(r,\hat{\mathbf{r}},\hat{\mathbf{u}}_i,\hat{\mathbf{u}}_j)$ **[36]**. It is, therefore, sufficient to have the expressions for $\partial V/\partial r$, $\vec{\nabla}_{\hat{\mathbf{r}}}V$ and $\vec{\nabla}_{\hat{\mathbf{u}}_i}V$ in order to obtain all the necessary forces and torques from them.

$$\frac{\partial V}{\partial r} = 24\varepsilon_0(2R^{-13} - R^{-7})/\sigma_0, \text{ if } r < r_l$$
$$= \frac{6(r_u-r)(r-r_l)\varepsilon_{wd}(\hat{\mathbf{r}},\hat{\mathbf{u}}_i,\hat{\mathbf{u}}_j)}{\text{range}^3}, \text{ if } r_l \leq r < r_u \quad (14)$$
$$= 0, \text{ otherwise.}$$

Now,

$$\vec{\nabla}_{\hat{\mathbf{r}}/\hat{\mathbf{u}}_i}V = -\frac{\partial V}{\partial r}\vec{\nabla}_{\hat{\mathbf{r}}/\hat{\mathbf{u}}_i}\sigma_{GB} - \vec{\nabla}_{\hat{\mathbf{r}}/\hat{\mathbf{u}}_i}\varepsilon_{wd}, \text{ if } r < r_l$$
$$= s(r)\vec{\nabla}_{\hat{\mathbf{r}}/\hat{\mathbf{u}}_i}\varepsilon_{wd}, \text{ if } r_l \leq r < r_u \quad (15)$$
$$= 0, \text{ otherwise.}$$

To get $\vec{\nabla}_{\hat{\mathbf{r}}/\hat{\mathbf{u}}_i}\sigma_{GB}$ and $\vec{\nabla}_{\hat{\mathbf{r}}/\hat{\mathbf{u}}_i}\varepsilon_{wd}$, note that defining

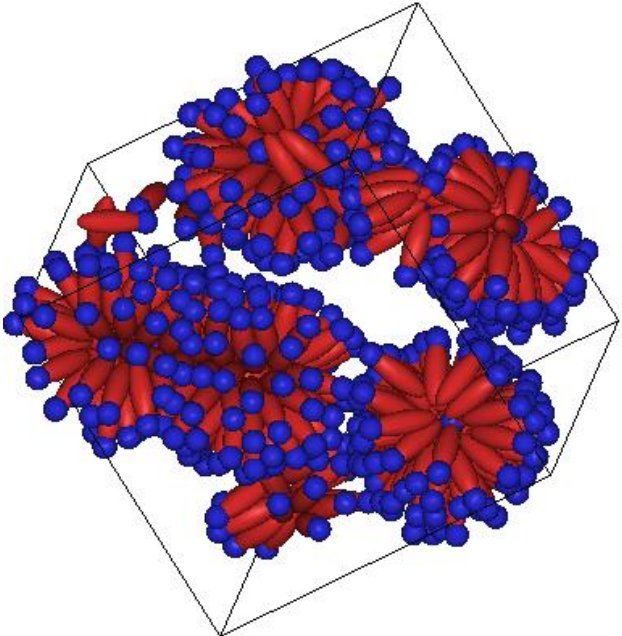

**FIG.12: Cylindrical and ellipsoidal micelles. Model parameters same as in FIG.11 except** $\varepsilon_e = 0.13$ **now. Dark spheres signify headgroups for directed surfactants.**

$$g(x) = \frac{1}{2}\left\{\frac{(\hat{\mathbf{r}}.\hat{\mathbf{u}}_i + \hat{\mathbf{r}}.\hat{\mathbf{u}}_j)^2}{1+x(\hat{\mathbf{u}}_i.\hat{\mathbf{u}}_j)} + \frac{(\hat{\mathbf{r}}.\hat{\mathbf{u}}_i - \hat{\mathbf{r}}.\hat{\mathbf{u}}_j)^2}{1-x(\hat{\mathbf{u}}_i.\hat{\mathbf{u}}_j)}\right\} \quad (16)$$

$$= \frac{(\hat{\mathbf{r}}.\hat{\mathbf{u}}_i)^2 + (\hat{\mathbf{r}}.\hat{\mathbf{u}}_j)^2 - 2x(\hat{\mathbf{r}}.\hat{\mathbf{u}}_i)(\hat{\mathbf{r}}.\hat{\mathbf{u}}_j)(\hat{\mathbf{u}}_i.\hat{\mathbf{u}}_j)}{1-x^2(\hat{\mathbf{u}}_i.\hat{\mathbf{u}}_j)^2}$$

we have $\sigma_{GB} = \sigma_0/\sqrt{1-\chi g(\chi)}$ and $\varepsilon_{GB} = 1-\chi' g(\chi')$. Hence, the knowledge of $\vec{\nabla}_{\hat{\mathbf{r}}/\hat{\mathbf{u}}_i} g(x)$ gives $\vec{\nabla}_{\hat{\mathbf{r}}/\hat{\mathbf{u}}_i}\sigma_{GB}$ and $\vec{\nabla}_{\hat{\mathbf{r}}/\hat{\mathbf{u}}_i}\varepsilon_{GB}$.

$$\vec{\nabla}_{\hat{\mathbf{r}}} g(x) = 2\frac{\zeta_j^i \hat{\mathbf{u}}_i + \zeta_i^j \hat{\mathbf{u}}_j}{1-x^2(\hat{\mathbf{u}}_i.\hat{\mathbf{u}}_j)^2}. \quad (17)$$

where $\zeta_j^i = \hat{\mathbf{r}}.\hat{\mathbf{u}}_i - x(\hat{\mathbf{r}}.\hat{\mathbf{u}}_j)(\hat{\mathbf{u}}_i.\hat{\mathbf{u}}_j)$. Again,

$$\vec{\nabla}_{\hat{\mathbf{u}}_i} g(x) = 2\frac{\zeta_j^i \hat{\mathbf{r}} + x[x(\hat{\mathbf{u}}_i.\hat{\mathbf{u}}_j)^2 g(x) - (\hat{\mathbf{r}}.\hat{\mathbf{u}}_i)(\hat{\mathbf{r}}.\hat{\mathbf{u}}_j)]\hat{\mathbf{u}}_j}{1-x^2(\hat{\mathbf{u}}_i.\hat{\mathbf{u}}_j)^2}. \quad (18)$$

Writing $[\nu_1(\hat{\mathbf{u}}_i \cdot \hat{\mathbf{u}}_j) + \nu_2(\hat{\mathbf{u}}_i \cdot \hat{\mathbf{r}} - \hat{\mathbf{u}}_j \cdot \hat{\mathbf{r}}) - \nu_3(\hat{\mathbf{u}}_i \cdot \hat{\mathbf{r}})(\hat{\mathbf{u}}_j \cdot \hat{\mathbf{r}}) + 1]$ as $\varepsilon'$, equation (5) becomes

$$\varepsilon_{wd} = \varepsilon_0 \varepsilon'^{\nu_0} \varepsilon_{GB}. \quad (19)$$

Hence,

$$\vec{\nabla}_{\hat{\mathbf{r}}/\hat{\mathbf{u}}_i}\varepsilon_{wd} = \nu_0 \frac{\varepsilon_{wd}}{\varepsilon'}\vec{\nabla}_{\hat{\mathbf{r}}/\hat{\mathbf{u}}_i}\varepsilon' + \varepsilon_0\varepsilon'^{\nu_0}\vec{\nabla}_{\hat{\mathbf{r}}/\hat{\mathbf{u}}_i}\varepsilon_{GB}. \quad (20)$$

The last piece needed is, therefore,

$$\vec{\nabla}_{\hat{\mathbf{r}}}\varepsilon' = \nu_2(\hat{\mathbf{u}}_i - \hat{\mathbf{u}}_j) - \nu_3[(\hat{\mathbf{u}}_j \cdot \hat{\mathbf{r}})\hat{\mathbf{u}}_i + (\hat{\mathbf{u}}_i \cdot \hat{\mathbf{r}})\hat{\mathbf{u}}_j], \text{ and} \quad (21)$$

$$\vec{\nabla}_{\hat{\mathbf{u}}_i}\varepsilon' = \nu_1\hat{\mathbf{u}}_j + [\nu_2 - \nu_3(\hat{\mathbf{u}}_j \cdot \hat{\mathbf{r}})]\hat{\mathbf{r}}. \quad (22)$$